\documentclass[twocolumn,amsmath,superscriptaddress,amssymb,pre]{revtex4}

\usepackage{amsmath,amssymb,latexsym,graphics,epsfig}
\usepackage{pstricks}
\usepackage{float}

\newpsobject{showgrid}{psgrid}{subgriddiv=1}

\newcommand \be {\begin{equation}}
\newcommand \ee {\end{equation}}
\newcommand \bea {\begin{eqnarray}}
\newcommand \eea {\end{eqnarray}}

\begin{document}

\title{Helices in the wake of precipitation fronts}

\author{Shibi Thomas}
\altaffiliation[Present address: ] {Department of Physics, University of Calicut, Kerala, India}
\email{shibithomas969@gmail.com}
\affiliation{Department of Theoretical Physics,  E\"{o}tv\"{o}s University, 1117 Budapest, Hungary}

\author{Istv\'an Lagzi}
\email{istvanlagzi@gmail.com}
\affiliation{Department of Physics, Budapest University of Technology and Economics, 1111 Budapest, Hungary}

\author{Ferenc Moln\'ar Jr.}
\email{molnaf@rpi.edu}
\affiliation{Department of Physics, Applied Physics, and Astronomy, Rensselaer Polytechnic Institute, Troy, New York 12180, USA}

\author{Zolt\'an R\'acz}
\email{racz@general.elte.hu}
\affiliation{Institute for Theoretical Physics - HAS,
  E\"otv\"os University, 1117 Budapest, Hungary}

\date{\today}

\begin{abstract}
A theoretical study of the emergence of helices in the wake of 
precipitation fronts is presented. The precipitation dynamics is 
described by the Cahn-Hilliard equation and the 
fronts are obtained by quenching the system into  a
linearly unstable state. Confining the process 
onto the surface of a cylinder and using the pulled-front 
formalism, our analytical calculations show
that there are front solutions that propagate into the 
unstable state and leave behind a helical structure. 
We find that helical patterns emerge only if the radius of the
cylinder $R$ is larger than a critical value $R>R_c$, in agreement
with recent experiments. 
\end{abstract}

\maketitle

\section{Introduction}
\label{Intro}

Chiral patterns have been the subject of a large number of studies in 
natural sciences and engineering, as well as in the artistic domain 
\cite{Su-doublehelix,Imai-chiral,Gao-Zincoxid}. 
The emergence of chirality at meso- and macro-scale is usually 
a complex process that may go along principally distinct routes. 
First, the chirality may be present in the microscopic 
building blocks and the symmetry is just transcribed to 
a higher level of spatial organization \cite{helicats}. 
Second, achiral microscopic entities may assemble into 
chiral objects provided the process takes place in a 
chiral medium \cite{chiral-fibers}. 
Finally, achiral microscopic units may self-organize 
into a chiral structure through symmetry breaking \cite{Sci-1982-Muller}.
 
Our interest is in the symmetry breaking route, and this work is
a follow-up to our recent studies \cite{Shibi2012,Shibi-Mat-Pack} in which 
helical precipitation patterns were observed 
in the wake of moving reaction-diffusion 
fronts. In our experiments, we saw no chirality 
in the precipitation blocks at the microscale \cite{Shibi-Mat-Pack} 
and, furthermore, the media, 
the precipitation dynamics, and the boundary conditions of the 
laboratory setup also lack chirality, thus we believe that the macroscopic 
patterns form through symmetry breaking.   
This view was also confirmed by simulations \cite{Shibi2012} 
that suggest that
the helices emerge from a complex interplay 
among the unstable precipitation modes, the motion of the  
reaction front, and the noise in the system. 

\begin{figure}[htb]
  \includegraphics[width=4cm]{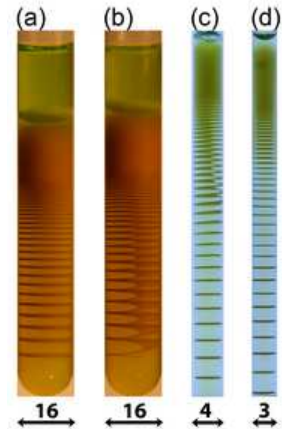}
\vspace{-10pt}
\caption{Liesegang experiments producing precipitation patterns in the
form of (a) bands or (b) helices under identical external experimental 
conditions in tubes of radius $R=8$ mm (the numbers below the tubes 
are their diameters in mm).  
(c) Example when the radius is smaller 
($R=2$ mm) and the helix  becomes 
unstable as its pitch increases.
(d) Only Liesegang bands form in tubes with radius $R\leq 1.5$ mm. 
\label{Fig:Liese}}
\end{figure}

Although the simulations correctly describe the 
trends observed in experiments, one would also like to make 
analytical advances in at least some aspects of the above problem. 
In experiments using Liesegang-type setups (Fig.\ref{Fig:Liese}),
we measured the probability of the emergence of helices 
as a function of control parameters such as the concentrations 
of the inner and outer electrolytes, the temperature, and the 
radius $R$ of the test tube. 
A simple but remarkable feature of the observations 
(reproduced in simulations) is that the probability approaches
zero at a critical radius $R_c$ below which no helical pattern 
forms. Our aim with this paper is to provide an explanation 
for the nonexistence of helical solutions below a critical radius 
$R_c$ using analytical calculations within the theoretical framework
of Cahn-Hilliard precipitation dynamics \cite{CahnHill1958}
which has been used successfully in simulations 
to interpret the experimental results \cite{Shibi2012}.  

It should be noted that theoretical results about the absence of 
helical patterns below $R_c$ have been 
derived earlier on phenomenological grounds  \cite{Polezhaev1991,Polezhaev1994}. Our work  
is based on similar logic in the sense 
that the conclusion is obtained by considering 
propagating helical waves evolving from an 
unstable state in precipitation dynamics. 
The differences lie in the use of a more transparent model 
of precipitation, and in the well defined approximation 
(pulled front formalism \cite{Saarloos2003}) 
used for the analytic derivation of the bound on $R_c$.

We describe the experimental background and the results
motivating our study in Sec.\ref{exp}, while the theoretical model 
and the pulled-front formalism 
are summarized in Sec.\ref{theory}. The theory is first 
applied (Sec.\ref{Lbands})
to the emergence of regular Liesegang patterns (bands parallel to
the front). Then helical solutions (bands tilted with respect to 
the front) are obtained (Sec.\ref{helix}), and 
the conditions for the existence of helical solutions
are derived (Sec.\ref{Rlarger}). 
We conclude with discussions of more
complex patterns and by reviewing the 
unsolved aspects of the problem (Sec.\ref{Concl}).

\section{Experiments}
\label{exp}

Precipitation patterns have captivated the imagination for a long time
\cite{Henisch1991} and systematic studies of 
the so-called Liesegang bands (Fig.1) have been going on 
for more than a century \cite{Liesegang}.
In a typical Liesegang type experiment, 
a gel column soaked with a chemical reactant
(called the inner electrolyte and denoted by $B$) is placed 
in a tube and another reactant 
(the outer electrolyte, denoted by $A$) is poured 
over the gel. The initial concentration of the outer electrolyte
is chosen to be much larger than that of the inner 
electrolyte ($a_0\gg b_0$), thus a diffusive front moves into the gel
where the reactions ($A+B\to \ldots \to C$) take place. For appropriate 
choice of reagents and initial concentrations, the final product $C$
emerges as a precipitate and the region of high concentrations of 
$C$ becomes visible as a pattern in the wake of the front. 

The simplest patterns are the much studied
Liesegang bands [see Fig.1(a)] which 
have been shown to obey a set of laws
governing the distance between the consecutive bands, 
the width of the bands, and their
time of appearance \cite{Henisch1991,MullerRoss2003,MathPack-1955,MatPack98,Liese-width1999,LieseRZ1999}.
There are, however, complex precipitation patterns 
displaying curiosities such as bandsplitting,
irregular banding, spirals, helices [see Fig.1(b,c)] and 
secondary- and revert patterns \cite{Sci-1982-Muller,Shibi2012,Henisch1991,Mathur-revert,Sultan-revert,Volfi2007,Mofi2008,Lagzi-Grzy} 
which are less readily explained. Frequently, they are just 
peculiarities of a given system, and some of them have
problems with reproducibility. Our experiments \cite{Shibi2012}, however, proved that
the emergence of helices is a robust phenomenon: They appear 
reproducibly with well defined probabilities
for a given range of experimental parameters.

In our experiments, described in more detail in 
\cite{Shibi2012,exp-Lagzi}, we used potassium chromate ($B\rm \equiv K_2CrO_4$) and copper chloride ($A\rm \equiv CuCl_2$) as
the inner and outer electrolytes, respectively. The 
solid precipitate emerged from the reaction 
${\rm Cu^{2+}+CrO_4^{2-}\to CuCrO_4\equiv }\, C$ which
took place in a 1\% agarose gel with the 
temperature kept constant ($T=22{\rm ^oC}$).
Below we display results for the following initial concentrations 
of the electrolytes: $\rm [Cu^{2+}]_0={\it a}_0=0.5 M$ 
and $\rm [ CrO_4^{2-}]_0={\it b}_0=0.01 M$. The experiments were 
carried out for a set of test-tube radii in the range 
$\rm 1.5 mm \le {\it R} \le 12.5 mm$, and an estimate of the probability $P_H$
of the emergence of helices was obtained from ten experiments for each
$R$ (Table \ref{table1}).
\begin{table}[htb!]
\caption{Probability of the emergence of helical pattern $P_H$ in 
experiments in which the test tube radius $R$ was the only parameter varied (see also Fig.1 in \cite{Shibi2012}).}   
\begin{ruledtabular}
\begin{tabular}{|c|c|c|c|c|c|c|c|c|c|c|c|}
$R$ (mm)&1.5&2&3&4&5&6&7&8&9&10&12.5\; \\[2pt]
\hline\noalign{}
$P_H$ &0&0.1&0.1&0.2&0.1&0.2&0.3&0.7&0.2&0.2&0.1 \\[0pt]
\end{tabular} 
\end{ruledtabular}  
\label{table1}
\end{table} 

As one can see from Table \ref{table1}, the probability has a maximum
around $R\approx 7-8$ mm, it decreases for large $R$ (due to the emergence of
more complex structures such as double helices or chaotic patterns) as well 
as for small $R$, and it goes to zero at $R_c\approx 1.5$ mm. Below,
we shall analytically derive the nonexistence of helices for $R<R_c$.

\section{Theory}
\label{theory}
Since the helical 
structures emerge at the macroscale and they can be viewed 
as slight variations of the usual Liesegang bands, 
we expect that they can also be analyzed within the 
framework of Cahn-Hilliard dynamics combined with a moving 
reaction front providing the precipitating material \cite{ModelB1999}. 
This approach has been successful in deriving the various laws 
describing the Liesegang bands, and it has helped in the understanding of 
how to control the bandspacing by external fields 
\cite{Guiding-field-07,Liese-2008}.

We shall actually further simplify the description, namely,  
the stage of the formation of the reaction product is replaced by 
an initial condition where the reaction product is homogeneously 
distributed with the concentration $c_0$. 
It is known that the diffusive reaction front leaves behind a 
homogeneous state of the reaction product \cite{MatPack98,LieseRZ1999} 
and helices usually form when a fast moving front prepares 
a relatively large region of the system 
in an unstable state \cite{Shibi2012}. 
We shall assume that this unstable state is the initial state
for the precipitation dynamics  studied by using 
the Cahn-Hilliard equation \cite{CahnHill1958,Halp-Hoh1977}
\be
\partial_t m=-\Delta (m-m^3+\Delta m)\, .
\label{CH-sc}
\ee 
Here the field $m$ is a shifted and rescaled concentration 
with $m=\pm 1$ corresponding to the 
high- and low-concentration equilibrium 
values, and ${\cal F}(m)=-m^2/2+m^4/4+(\nabla m)^2/2$ 
is the free-energy 
density underlying the drive towards equilibrium.
The coefficients in Eq.\eqref{CH-sc} are set to unity by
choosing the length, time, and concentration scales 
appropriately. 

\begin{figure}[htb]
  \includegraphics[width=8cm]{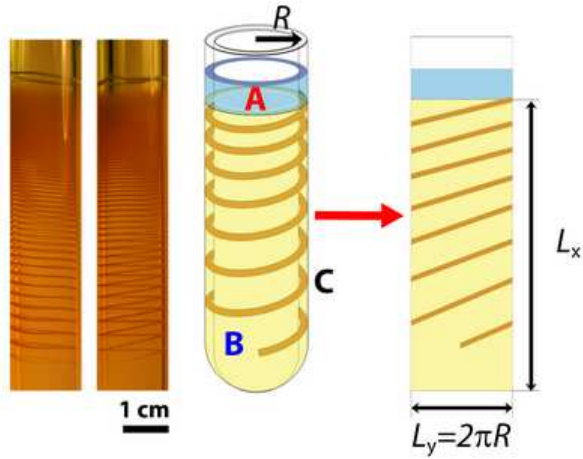}
\caption{Liesegang type experiments with the precipitation
patterns forming in the gel placed in between two tubes of 
nearly equal radius (Liesegang rings in the leftmost tube and a
helix in the next one). On the right, a 
schematic drawing is displayed showing the transformation of 
the thin layer of gel in the tube-in-tube experiment 
into a two-dimensional strip. 
\label{F:helix-plain}}
\end{figure}

Equation \eqref{CH-sc} is considered in a two-dimensional strip 
corresponding to the tube-in-tube experiments 
\cite{Shibi2012} where the helices emerge in a thin layer 
of gel in between two tubes of nearly equal radius. The 
cylinder can be cut and opened into a strip as shown in 
Fig.\ref{F:helix-plain} with the transformation implying
that we have periodic boundary conditions across the strip.  
Initially, a homogeneous state $m(x,y,t=0)=m_0$
is prepared that is linearly unstable, i.e. the concentration 
is within the spinodal decomposition range, $|m_0|<1/\sqrt{3}$.
 
Such an initial state is stationary, and so we add a small 
local perturbation $m_0\to m_0+\delta m(x,y,0)$ with $
\delta m(x,y,0)$ restricted to the region $x\approx 0$, $0<y<L_y$.
The perturbation develops into two precipitation fronts moving 
in the $\pm x$ direction and the question we pose is about the nature of
patterns left behind the fronts. More precisely, we ask if a 
helix that is a striped pattern tilted with 
respect to the propagation direction is present among the solutions. 
In order to answer this question, we
assume that the dynamics in the front region (where 
$\delta m =m-m_0\ll 1$) can be described by the 
linearized theory i.e. we assume that the front belongs
to the pulled-front family \cite{Saarloos2003}. The theory 
of pulled fronts has been employed successfully to the $d=1$ 
Cahn-Hilliard equation \cite{Liu-Gold-89,Burki2007,Krekhov2009}. 
Below, we repeat (in a 
non-rigorous form) the main steps of the theory 
in order to clearly outline the assumptions needed for the 
generalization to the strip geometry of interest.

The first step in the theory of pulled fronts is 
the linearization of the equation in question, i.e. 
we write $m=m_0+\delta m$ and obtain from Eq.\eqref{CH-sc}
\be
\partial_t \delta m=-\Delta (a +\Delta) \delta m
\label{CH-lin1}
\ee 
where $a=1-3m_0^2$ is a measure of the distance of the 
initial state from the spinodal ($a=0$). Equation \eqref{CH-lin1}
can be solved by Fourier transformation
\be 
\delta m(x,y,t)=\frac{1}
{\sqrt{2\pi L_y}}\sum_{k_y}\int\limits_{-\infty}^\infty dk_x e^{i(k_xx+k_yy)}m_{\bf k}(t)
\label{Fourier1}
\ee
where ${\bf k}=(k_x,k_y)$ and, due to the periodic boundary 
conditions in the $y$ direction, we have $k_y=2\pi n/L_y$ with 
$n=0,\pm 1,...,\pm(L_y/2-1),L_y/2$. The Fourier components of the perturbation 
$m_{\bf k}$ evolve independently
\be
m_{\bf k}(t)=e^{\omega_{\bf k}t}m_{\bf k}^0\, ,
\label{CH-lin-sol}
\ee  
with $\omega_{\bf k}$ obtained by substituting \eqref{CH-lin-sol} 
into \eqref{CH-lin1} 
\be
\omega_k=ak^2-k^4\, ,
\label{disp}
\ee
where $k^2=k_x^2+k_y^2$.

In order to evaluate \eqref{Fourier1}, the 
initial amplitudes $m_{\bf k}^0$ need to be specified. Since the
initial perturbation is restricted to the $x\approx 0$ region, 
$m_{\bf k}^0$ is practically independent of $k_x$ and so 
$m_{\bf k}^0\approx m_{k_y}^0$. Thus we can write 
\eqref{Fourier1} in the form
\be 
\delta m(x,y,t)\approx \frac{1}{\sqrt{2\pi L_y}}\sum_{k_y}e^{ik_yy}m_{k_y}^0
\hspace{-3pt}\int\limits_{-\infty}^\infty dk_x e^{{ik_xx}+\omega_kt} .
\label{Fourier2}
\ee
We shall now analyze the above expression term by term. 

\section{Liesegang-type patterns}
\label{Lbands}

It is clear 
that the $k_y=0$ term describes a one-dimensional pattern that is 
homogeneous in the $y$ direction. This brings us back to the 
one-dimensional case where, in the frame moving with the velocity $v_0$ 
of the front, we have
\be 
\delta m(v_0t+\xi,0,t)\sim 
\int\limits_{-\infty}^\infty dk_x e^{ik_x\xi}\exp{({ik_xv_0}+\omega_{k_x})t}\, .
\label{dm-front}
\ee
A saddle-point evaluation of the $t\to\infty$
asymptote of the above integral, together with the requirement 
that $\delta m$ remains finite in the front region, 
leads to the basic equations of the theory of pulled fronts 
\cite{Saarloos2003}
\be
iv_0+\left.\frac{d\omega_{k_x}}{dk_x}\right |_{k_x^*}=0\quad ;\quad {\rm Re}\,(ik_x^*v_0+\omega_{k_x^*})=0 \; .
\label{x-triple}
\ee 
The above equations determine $v_0$ and $k_x^*=p_0+iq_0$ 
with $p_0$ and 
$q_0$ related to the characteristic wavenumber of the pattern in the 
comoving frame, and to
the steepness of exponential decay of the front profile
\bea
&&\delta m(v_0t+\xi,0,t) \nonumber \\
&&\sim \exp{\left[-q_0\xi+ip_0\xi +i(p_0v_0+{\rm Im}\, \omega_{k_x^*})t\right]} 
\, .
\label{x-front} 
\eea
The values of $v_0$, $p_0$, and $q_0$ can be easily calculated from 
\eqref{x-triple} and one obtains \cite{Saarloos2003}
\bea
p_0&=&\frac{1}{2}\sqrt{\frac{\sqrt{7}+3}{2}}\,a^{1/2}\approx 0.840\,a^{1/2}\, ,\label{homo1}\\
q_0&=&\frac{1}{2}\sqrt{\frac{\sqrt{7}-1}{6}}\,a^{1/2}\label{homo2}\approx 0.262\,a^{1/2}\, ,\\
v_0&=&\frac{2}{3}\sqrt{\frac{7\sqrt{7}+17}{6}}\,a^{3/2}\approx 1.622\,a^{3/2}\,.
\label{homo3}
\eea
As we can see from the above results [Eqs.(\ref{homo1}--\ref{homo3})], 
the spinodal ($a=0$) can be viewed as a 
critical point. Indeed, when approaching the spinodal ($a\to 0$), 
the characteristic length-scales 
($\ell\sim 1/p_0\sim 1/q_0\sim a^{-1/2}$) and the characteristic 
time-scale ($\tau \sim \ell/v_0\sim a^{-2}$) diverge as in mean-field 
theories of critical phenomena.

The wavenumber of the pattern 
in the laboratory frame, $p^{st}_0$, is obtained by noting that, 
apart from the late stage coarsening process,  
the pattern becomes stationary in the laboratory frame. Thus $p^{st}_0$ 
is calculated by equating the frequency of precipitation bands leaving 
from the front region ($v_0p^{st}_0/2\pi$) to the frequency of the arrival of the perturbation maxima in the comoving 
frame $(p_0v_0+{\rm Im}\, \omega_{k_x^*})/2\pi$. As a result we 
find
\be 
p^{st}_0=p_0+\frac{{\rm Im}\, \omega_{k_x^*}}{v_0}=\frac{\sqrt{7}+1}{4}p_0\approx 0.766\,a^{1/2}\, .
\label{lab_fr1}
\ee 
Accordingly, the wavelength of the pattern (spacing of the 
precipitation bands) before the possible coarsening may take place 
is given by 
\be
\lambda_0^{st}=\frac{2\pi}{p^{st}_0}=\frac{16\pi}{3}\sqrt{\frac{19-7\sqrt{7}}{2}}a^{-1/2}\approx 8.206 a^{-1/2}.
\label{bandspace}
\ee

In some forms, the results embodied in 
Eqs.(\ref{homo1}--\ref{lab_fr1}) have been derived in Refs.
\cite{Saarloos2003,Liu-Gold-89,Burki2007,Guiding-field-07,Krekhov2009,Foard-Wagner-09,Foard-Wagner-11,Kopf-2012} where the front velocity and the 
characteristic length were calculated in various quench-related problems. 
The results were also used 
to describe enslaved phase separation dynamics \cite{Guiding-field-07,Foard-Wagner-09,Foard-Wagner-11}
where the velocity of the front was slowly changing 
as prescribed by external fields. The logic in the 
present paper is similar to that used in the latter works. 
Namely, the wavelength of the pattern is identified as a changing
local wavelength related to the velocity of the front and 
frozen in the wake of the front \cite{Guiding-field-07}. 
In case of Liesegang bands, this means that
the front moves diffusively and slows down and, consequently, 
the distance between consecutive bands increases 
yielding the observed geometric series for the band 
positions \cite{Foard-Wagner-11}.

\section{Single-helix pattern}
\label{helix}

\begin{figure}[htb]
  \includegraphics[width=8cm]{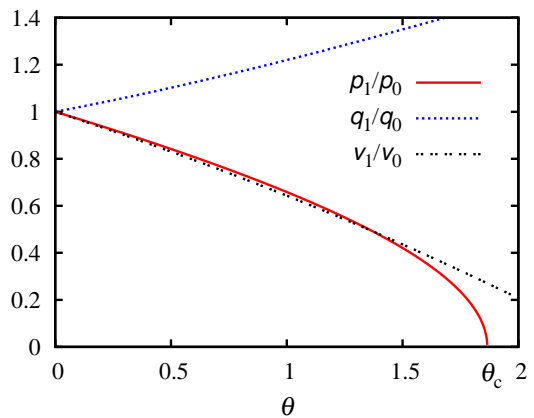}
\vspace{-10pt}
\caption{Characteristic wavenumbers $p_1$ and $q_1$ and the velocity 
$v_1$ of
the propagating single helix pattern scaled by the corresponding 
homogeneous values $p_0$, $q_0$ and $v_0$. The parameter $\theta$ 
is inversely proportional to the square of the diameter of the 
tube \eqref{theta}. As can be seen, propagating helix solutions exist
only for $\theta\leq \theta_c$ i.e. only for tubes with large enough 
diameter.  
\label{Fig:theta}}
\end{figure}

Next, we consider the case when 
the longest wave-length ($n=1$) transverse mode is excited only, i.e. 
the only nonzero amplitude in \eqref{Fourier2} is related to 
the mode $k_y=2\pi/L_y\equiv\kappa_1$. Thus, 
the initial perturbation takes the form
\be 
\delta m(x,y,t=0)\sim e^{i\kappa_1 y}
\int\limits_{-\infty}^\infty dk_x e^{{ik_xx}+\omega_kt}\, .
\label{init2}
\ee
As in the $d=1$ case, the above expression is analyzed
in the frame moving with the velocity $v_1$ of the front
\be
\delta m(v_1t+\xi,y,t=0)
\sim e^{i\kappa_1 y}
\int\limits_{-\infty}^\infty dk_x e^{ik_x\xi+{(ik_xv_1}+\omega_k)t}\, ,
\label{dm-front-2}
\ee  
where one expects a new value for the velocity front $v_1$ since 
$\omega_k$ now depends on $k^2= k_x^2+\kappa_1^2$.  The equations 
to be solved remain the same \eqref{x-triple} with $\omega_{k_x}$
replaced by $\omega_k$. Let us denote the solution of the
equations by $k_x^*=p_1+iq_1$ where $p_1$ and $q_1$ depend 
not only on $a$ but also on $\kappa_1$. Then the perturbation 
in the comoving frame takes the form
\bea
&&\delta m(v_1t+\xi ,y,t=0)\nonumber \\
&&\sim \exp{[-q_1\xi+i(\kappa_1 y+
p_1\xi+{\rm Im}\omega_{k^*}t)]}\, .
\label{dm-front-3}
\eea 
with 
\bea
p_1&=&\sqrt{\frac{\sqrt{(1-\theta)^2+6}+3(1-\theta)}{\sqrt{7}+3}}\,p_0\label{heli1}\\
q_1&=&\sqrt{\frac{\sqrt{(1-\theta)^2+6}-(1-\theta)}{\sqrt{7}-1}}\,q_0\label{heli2}\\
v_1&=&\frac{{\rm Re} \, \omega_{k^*}}{{\rm Im}\, k^*}\\ &=&\sqrt{\frac{[(1-\theta)^2+6]^{{3}/{2}}-(1-\theta)^3 +18(1-\theta)}{7\sqrt{7}+17}} v_0\nonumber
\label{heli3}
\eea 
where the parameter $\theta$ is related to the width of the strip 
$L_y$ and the radius of the cylinder $(L_y=2\pi R)$ through
\be
\theta =2\frac{\kappa_1^2}{a}=2\frac{(2\pi)^2}{L_y^2a}=\frac{2}{R^2a}\, .
\label{theta}
\ee
It can be easily verified that, in the infinite width limit 
($R\to\infty$ or $\theta \to 0$), we recover the parameters 
of the solution homogeneous
in the $y$ direction [Eqs.\eqref{homo1}--\eqref{homo3}]. The scaled
variables $p_1/p_0$, $q_1/q_0$ and $v_1/v_0$ are displayed in
Fig. \ref{Fig:theta}.

According to Eq.\eqref{dm-front-2}, the solution 
(\ref{heli1})--(\ref{heli3}) is a 
wave of tilted precipitation bands propagating in the comoving frame. 
Due to the periodic boundary conditions in the transverse direction,
the corresponding pattern on the cylindrical surface is a propagating 
helix. The pitch of the helix in the comoving frame is given by 
$\lambda_1=2\pi/p_1$ and, to obtain the pitch in the 
laboratory frame, $\lambda_1^{st}=2\pi/p_1^{st}$,
we have to use again the stationarity of the pattern in 
the laboratory 
$p^{st}_1=p_1+{\rm Im}\, \omega_{k^*}/v_1 
\label{lab_frame1}$, 
resulting in
\be
\lambda_1^{st}=\frac{2\pi}{p^{st}_1}=\frac{16\pi\sqrt{2}}{3}
\frac{\sqrt{(1-\theta)^2+6}+2(1-\theta)}{[\sqrt{(1-\theta)^2+6}+3(1-\theta)]^{3/2}}a^{-1/2}.
\label{helixpitch}
\ee
We have thus found helix solutions with well defined propagation 
velocity and pitch determined by 
the parameter $a$ of the Cahn-Hilliard equation and by the 
width $L_y=2\pi R$ of the system.

\section{Existence and relevance of the helix solutions}
\label{Rlarger}

Examining $p_1$, $q_1$ and $v_1$
(\ref{heli1}--\ref{heli3}) reveals that the 
propagating helix solutions exist
only for sufficiently small values of $\theta$. Indeed, as $\theta$ is 
increased from zero, 
the expressions under the square root in $p_1$ and $v_1$ 
become negative thus contradicting our assumptions 
that $p_1$, $q_1$ and $v_1$ are real. The smallest critical value 
of $\theta$ is obtained from the equation $p_1(\theta_c)=0$ with the result 
\be
\theta_c=1+\sqrt{3}/2\approx 1.866\, .
\label{crittheta} 
\ee
Thus, we arrive at our main result, namely, 
helix solutions exist only if $\theta \le \theta_c$. In terms of
the radius $R$ of the test tubes, this inequality means that helices 
may emerge in a test tube only if $R$ exceeds a critical value
\be
R\ge R_c=\sqrt{\frac{2}{\theta_c a}}\approx 1.035\,a^{-1/2}\,.
\label{Rc1}
\ee 
Estimating $R_c$ for a given system runs into the problem that
$R_c$ is measured in unknown units since 
the details of mapping of the system onto the Cahn-Hilliard dynamics
are usually lacking. We can go around this problem by
obtaining the lengthscale from the results for the 
spacing of bands formed parallel with the front ($\theta =0$). The 
remarkable feature of this case is that the band spacing 
$\lambda_0^{st}$ is independent of $R$. Thus,
using Eq.\eqref{bandspace}, we can write the inequality 
\eqref{Rc1} in a simple form  
\be
R\ge R_c\approx 0.126\,\lambda_0^{st}\,.
\label{Rc2}
\ee 
To a good approximation, the above inequality means that 
helices can form 
if the diameter of the test tube is larger than 1/4 of the 
band spacing in identical experiments where bands 
were formed.

We can try now to carry out a straightforward comparison with the 
experiments and examine whether the inequality (25) is violated in 
cases in which no helices are observed. In the experiments, the diameters
of the tubes $D=2R$ range in the interval 25 mm $\ge D \ge$ 3 mm. 
The radius is fixed for a given 
experiment, in contrast to the bandspacing $\lambda_0^{st}$ (or 
to the pitch $\lambda_1^{st}\approx \lambda_0^{st}$ of the helices) 
which changes within each pattern. In order to look for 
violation of the inequality \eqref{Rc2}, we took the largest 
values of $\lambda_0^{st}$ or  $\lambda_1^{st}$ for 
each pattern and determined the corresponding smallest 
possible ratios $D/\lambda_0^{st} = u$ or $D/\lambda_1^{st} = u$. 
For the experiments displayed in Fig.\ref{Fig:Liese}(a-d), 
we found $u_a >3.8$, $u_b>3.6$, $u_c>1.6$ and $u_d>1.2$. 
Thus the smallest $u$-s are 
significantly larger than 1/4 and the inequality (25) is not 
violated even in 
cases when helices are not formed. The conclusion remains the same 
if all the smallest $u$-s are calculated for 
the experiments and simulations studied in \cite{Shibi2012}. 
Clearly, the comparison with experiments works only at the 
qualitative level, namely, decreasing $R$ leads to the violation 
of the inequality and to the absence of helices, and this is 
in agreement with the observations.

We should emphasize that it is not surprising that we see only 
qualitative agreement. One should remember that 
the results, including the inequality \eqref{Rc2}, 
apply to propagating precipitation fronts. 
Thus, extending them to diffusive fronts 
such as the ones producing Liesegang bands or helices involves additional 
assumptions. First, the local velocity of the front is assumed to
be identical to the linearly selected pulled front velocity. 
Second, the local wavelength of the pattern (bandspacing or pitch) 
emerging in the wake of 
the front is assumed to be frozen without any further coarsening.
Using these two assumptions seems to work well when interpreting 
patterns formed in enslaved phase separation processes 
\cite{Guiding-field-07,Foard-Wagner-09,Foard-Wagner-11}, thus they 
can be viewed as reasonable assumptions. Naturally, one should 
suspect that while the mapping of the propagating front onto 
a diffusive one may leave the inequality \eqref{Rc2} qualitatively valid, 
the constant $u$ in $2R>u\lambda_0^{st}$ \eqref{Rc2} will 
be affected.

Unfortunately, there are additional problems when comparing 
the inequality \eqref{Rc2} with Liesegang-type 
patterns. The pattern often evolves from a homogeneous precipitate 
called plug (see the upper part of the precipitation in 
Fig.\ref{Fig:Liese}) with the initial 
bandspacing $\lambda_0^{st}$ being small and not always well resolved. 
Furthermore, the band-spacing grows exponentially, thus  
the $2R>\lambda_0^{st}/4$ rule \eqref{Rc2} should always be violated 
for long enough tubes. Of course, the experimental tubes are finite
and, as can be seen in the example shown in Fig.\ref{Fig:Liese}(a), 
the $2R>\lambda_0^{st}/4$ rule is satisfied throughout the system.
Thus, in this case one expects that 
there is no problem observing helical patterns, as indeed is 
the case [Fig.\ref{Fig:Liese}(b)]. We have also seen examples when
the band spacing is larger and 
the helical pattern becomes unstable as its pitch 
increases [Fig.\ref{Fig:Liese}(c)], or when no helix forms 
at all [Fig.\ref{Fig:Liese}(d)].
Whether this is the result of violating
the inequality  $D/\lambda >u$ with an effective 
(and presently unknown) $u$  remains an open question
since coarsening and other nonlinear effects 
may always have unexpected effects on the stability of helices. 

The trends  in the experimental observations 
and in the related simulations \cite{Shibi2012} are, however, 
in agreement with the analytical result 
\eqref{Rc2}. Thus we feel that the assumptions required to  
extend the results of the pulled-front theory to diffusive 
fronts are valid and, consequently the inequality $D/\lambda >u$ 
is a relevant condition for helix formation 
in the wake of diffusive reaction fronts.

\section{Discussions}
\label{Concl}

One can easily verify that in addition to the propagating helix 
solutions (\ref{heli1})--(\ref{heli3}), one can also find \hbox{double-,} triple-, and multiple-helix solutions. Indeed, one just repeats 
the helix calculation with $\kappa_1$ replaced by 
$\kappa_n=n^2\kappa_1$, where $n$ is the multiplicity of the helix.
One can also verify, by noting that the effective $\theta$ for 
a helix with multiplicity $n$ is $n^2\theta$, 
that larger multiplicity results in smaller velocity and 
larger pitch. Furthermore, it also follows then 
that $\theta_c(n)=\theta_c/n^2$ and,
consequently, the larger the multiplicity, the larger 
the threshold is for the tube diameter for the multiple-helix solution 
to exist.    

In our experiments and simulations, we did observe single helices 
with large probability. Double helices had significant probabilities
in large systems and at high noise levels (in the simulations). 
Although triple helices were also seen, their probability was 
negligible (could not be measured within the number of experiments
and simulations carried out). Thus the modes we have been 
investigating do appear in the system and the outcome of their 
competitions seems to be determining the patterns emerging. 

There are, of course, a number of problems to 
solve before the mode competition in the helix formation
is fully understood. The stability of the helix solutions 
is clearly a relevant issue. One may expect that the 
helices are unstable in the linear regime (their velocity, 
e.g., is smaller than the velocity of the $\theta =0$ band 
solutions). As the experiments and simulations 
suggest, however, the helices are stabilized by the nonlinear 
effects. Thus, investigating the lifetime of the helices 
in the linear regime (as compared to the time the front moves 
the distance of the band spacing) may give an indication of how 
to start a calculation of $P_H$.  

Clearly, the effect of noise is also important since
the probability of the emergence of helices $P_H$ is negligible
at small noise and it becomes of the order of 0.5 for appropriate 
noise amplitude. The origin of noise is not entirely clear. It may 
come as an initial-state noise left behind the fast moving
reaction front. Another view (taken in \cite{Shibi2012}) 
is that the inhomogeneities 
produced by the front are negligible, and the noise present in the
precipitation process is the relevant effect. Finding 
the origin of noise would be essential in deciding whether
the emergence of chirality is due to the initial-state effects
or it is a symmetry breaking occurring in the course of the 
precipitation dynamics.

Another unclarified aspect of the problem is related to the 
boundary conditions. In the present paper, we assumed that the
front is already at infinity, and we have to care only about the 
transverse boundary conditions (which are obviously periodic). 
In reality, the front is diffusive and, although it may move 
fast at the beginning, it can be seen to interact with the
developing pattern \cite{Shibi2012}. Thus, the front may be 
relevant in the delicate interplay of the unstable modes and so, 
even if we would consider the front
as a stationary wall, the boundary condition on it is highly 
nontrivial and needs to be explored. 

In summary, we used analytical methods to 
understand a spatial constraint ($R>R_c$) in the formation of 
helices. Along the way, we also found that while 
the helices (and helicoids) are simple geometric 
objects, their formation through precipitation processes is
a rather complex and intriguing problem, and much 
remains to be understood.

\begin{acknowledgements}
The authors acknowledge the financial support of the Hungarian Research Found (OTKA K104666 and NK100296). F. M. has been partially supported by the NSF through Grant No. DEB-0918413.
\end{acknowledgements}

\end{document}